\documentclass[prl,showpacs,amsmath,amssymb,twocolumn,superscriptaddress]{revtex4-1}
\usepackage{bm,multirow}
\usepackage{lipsum,capt-of}
\usepackage[usenames]{color}
\newcommand \beq{\begin{eqnarray}}
\newcommand \eeq{\end{eqnarray}}
\def\simge{\mathrel{%
       \rlap{\raise 0.511ex \hbox{$>$}}{\lower 0.511ex \hbox{$\sim$}}}}
\def\simle{\mathrel{\rlap{\raise 0.511ex \hbox{$<$}}{\lower 0.511ex \hbox{$\sim$}}}}
\usepackage{graphics}
\usepackage[dvips]{graphicx}
\usepackage{graphicx}

\newcommand{\kk}{\mathbf{k}}

\newcommand{\rr}{\mathbf{r}}
\newcommand{\QQ}{\mathbf{Q}}
\newcommand{\KK}{\mathbf{K}}

\newcommand{\be}{\begin{equation}}
\newcommand{\ee}{\end{equation}}
\newcommand{\bea}{\begin{eqnarray}}
\newcommand{\eea}{\end{eqnarray}}
\newcommand{\ba}{\begin{align}}
\newcommand{\ea}{\end{align}}

\newcommand{\rme}{{\rm e}}
\newcommand{\rmi}{{\rm i}}

\usepackage[usenames,dvipsnames]{xcolor}
\usepackage[colorlinks=true,citecolor=Cerulean,linkcolor=RubineRed,urlcolor=Cerulean]{hyperref}

\usepackage[normalem]{ulem}
\usepackage{soul,xcolor}

\begin{document}
\title{
Superfluid Density of Neutrons in the Inner Crust of Neutron Stars:\\ New Life for Pulsar Glitch Models}
\author{Gentaro Watanabe}
\affiliation{Department of Physics and Zhejiang Institute of Modern Physics, Zhejiang University, Hangzhou, Zhejiang 310027, China}

\author{C. J. Pethick}
\affiliation{The Niels Bohr International Academy, The Niels Bohr Institute, University of Copenhagen, Blegdamsvej 17, DK-2100 Copenhagen \O, Denmark}
\affiliation{NORDITA, KTH Royal Institute of Technology and Stockholm University, Roslagstullsbacken 23, SE-106 91 Stockholm, Sweden}

\date{\today}

\begin{abstract}
Calculations of the effects of band structure on the neutron superfluid density in the crust of neutron stars made under the assumption that the effects of pairing are small {[}N. Chamel, Phys. Rev. C {\bf 85}, 035801 (2012){]} lead to moments of inertia of superfluid neutrons so small that the crust alone is insufficient to account for the magnitude of neutron star glitches.   Inspired by earlier work on ultracold atomic gases in an optical lattice, we investigate fermions with attractive interactions in a periodic lattice in the mean-field approximation.  The effects of band structure are suppressed when the pairing gap is of order or greater than the strength of the lattice potential.   By applying the results to the inner crust of neutron stars, we conclude that the reduction of the neutron superfluid density is considerably less than previously estimated and, consequently, it is premature to rule out models of glitches based on neutron superfluidity in the crust.  
\end{abstract}

\pacs{26.60.Gj, 21.60.Jz, 97.60.Gb, 03.75.Ss}

\maketitle

 An important problem in the physics of neutron stars is how to understand the sudden increases in the rotational frequencies, so-called glitches, that occur in many pulsars. One of the most promising models for understanding them is the sudden locking together of the interstitial neutron superfluid in the inner crust to the lattice of nuclei \cite{Haskell,GeneralRef}. 
An important quantity in the theory of glitches is the neutron superfluid density, which also enters in calculations of the frequencies of collective \mbox{modes \cite{Sauls}.}   On physical grounds one might expect the neutron superfluid density to be comparable to the density of neutrons between nuclei, which is what one finds in hydrodynamic and related approaches that do not take into account the band structure of the fermionic excitations that make up the neutron pairs \cite{hydro}.  An important question is whether neutron band structure due to the periodic lattice of nuclei can change this result significantly.  Theoretically, determining the neutron superfluid density is a difficult computational problem because neutrons can occupy as many as $\sim 500$ bands and it is necessary to take into account both band structure and the pairing interaction.  Chamel \cite{Chamel2012} performed calculations of the band structure of neutrons in the inner crust using a lattice potential obtained from microscopic calculations of the structure of nuclei in the crust \cite{Lattice_potentials}.   These calculations did not take pairing into account explicitly but should be a good approximation if the effects of pairing are weak.  They predict the neutron superfluid density to be a factor of 10 or more less than the naive estimate for a significant range of densities.  With such low values of the superfluid density, models of glitches based on weak coupling between a neutron superfluid and the lattice of nuclei \cite{Haskell} run into serious difficulties, as argued in Refs.~\cite{andersson, Chamel2013,Delsate}. 

 Here, we first present simple arguments to show that, in a fermionic superfluid, a potential  is less effective in scattering quasiparticles close to the Fermi energy than it is in the normal state.  We then solve a model of a fermionic superfluid with attractive interactions in a periodic potential that has previously been employed to treat the somewhat different problem of atomic Fermi gases with resonant interactions in an optical lattice, a periodic potential created by a standing electromagnetic wave \cite{Watanabe2008}.  These calculations demonstrate that band structure has a much smaller effect on the superfluid density when the pairing gap is larger than the strength of the lattice potential.   We shall then apply these results to the inner crust of a neutron star.   For the conditions under which the large reductions of the superfluid density were predicted in Ref.\,\cite{Chamel2012}, we find that for the majority of reciprocal lattice vectors, the strength of the lattice potential is less than the pairing gap and therefore they have little effect on the superfluid density.  When pairing is included, we estimate the reduction of the superfluid density to be some tens of a percent, rather than more than an order of magnitude, as in Ref.\,\cite{Chamel2012}.  We conclude that the moment of inertia of the neutron superfluid in the crust is large enough that glitch models based on the superfluid neutrons in the inner crust cannot be ruled out.

\textit{Simple example.---}
Consider a spin-independent potential described by the Hamiltonian
\be
H^{\rm int}=\sum_{\kk,\kk',\sigma} V(\kk-\kk')  a_{\kk,\sigma}^\dagger a_{\kk',\sigma}.
\label{scatpot}
\ee
Here, $a_{\kk,\sigma}^\dagger$ and $a_{\kk,\sigma}$ are creation and annihilation operators for fermionic particles with momentum $\kk$ {\cite{hbar} and spin projection $\sigma=\pm1$ in units of $\hbar/2$. They are related to the creation and annihilation operators $\alpha_{\kk,\sigma}^\dagger$ and $\alpha_{\kk,\sigma}$ for BCS quasiparticles in an S-wave fermionic superfluid by the expressions
\be
a_{\kk,\sigma}^\dagger=u_k \alpha_{\kk,\sigma}^\dagger+\sigma v_k \alpha_{-\kk,-\sigma}
\label{annihilation}
\ee
and its Hermitian conjugate.  In the absence of superfluid flow, we may take the coherence factors $u_k$ and $v_k$ to be real and positive, and
\be
u_k^2=\frac{1}{2} \left(1+\frac{\xi_k}{E_k}\right) \,\,\,{\rm and} \,\,\,v_k^2=\frac12 \left( 1-\frac{\xi_k}{E_k}\right),
\label{cohfactors}
\ee
where $E_k=\sqrt{\xi_k^2+\Delta^2}$ is the BCS quasiparticle energy.  Here, $\Delta$ is the pairing gap and $\xi_k=k^2/2m-\mu$, where $m$ is the particle mass and $\mu$ the chemical potential.  From Eqs.~(\ref{scatpot}) and (\ref{annihilation}), 
one sees that the amplitude to scatter an excitation with momentum $\kk$ and spin $\sigma$ to a state with momentum $\kk'$ and the same spin is
\be
\langle \kk' \sigma |H^{\rm int}|\kk \sigma\rangle =(u_ku_{k'}-v_kv_{k'})V(\kk-\kk'),
\ee
a result familiar in the context of metallic superconductors \cite{Tinkham}.
For excitations at the Fermi surface ($k=k'=\sqrt{2m\mu}$), $u_k=v_k =1/\sqrt{2}$ and therefore the scattering amplitude vanishes.  Physically, this is a consequence of the fact that excitations at the Fermi surface are superpositions of particles and holes with equal probabilities and the net scattering amplitude vanishes because the interaction potentials for particles and holes are equal and opposite.  This reduction of the effect of a scattering potential will result in band structure effects being suppressed by pairing.

\textit{Almost free-particle approximation with pairing.{---}}
The reduction of the scattering of quasiparticles at the Fermi surface reflects itself in the excitation spectrum.  We extend the standard model of band structure in a weak periodic potential, in which one takes into account the mixing of two single-particle states by the potential. Pairing is taken into account by including, in addition to the particle states  $\kk ,\sigma$ and $\kk' ,\sigma$ as is usually done, the corresponding hole states $-\kk, -\sigma$ and $-\kk', -\sigma$.  Thus, the eigenvalue problem becomes a $4 \times 4$ one, rather than the familiar $2 \times 2$ one when pairing is absent. When $\kk$ and $\kk'$ lie on the Fermi surface, one finds that the energy of an excitation measured with respect to the chemical potential is $\pm \sqrt{|\Delta|^2+V(\kk-\kk')^2}$, where $\Delta$ is the pairing matrix element.  In the absence of pairing this reduces to $\pm|V(\kk-\kk')|$, the result for the almost-free-particle model,  and in the absence of a periodic lattice, to $\pm|\Delta|$, the standard BCS result.  Of particular interest for the present problem is the fact that when $|\Delta|$ is large compared with $|V(\kk-\kk')|$, the changes in the spectrum due to the periodic potential are of order $|V(\kk-\kk')|^2/|\Delta|$, which is  $\sim |V(\kk-\kk')|/|\Delta|$ times their value $|V(\kk-\kk')|$ when pairing is absent.

\textit{Mean-field approach.{---}}
The primary aim of this Letter is to explore the competition between band structure and pairing.  We shall consider neutrons in a periodic external potential  $V_{\rm ext}(\rr)$; this potential represents the interaction of a neutron with the protons, which reside in nuclei, and other neutrons. In addition, we shall take into account the effects of the pairing interaction, which was neglected in Ref.\ \cite{Chamel2012}.   We shall adopt a mean-field approximation, which is the Bogoliubov--de Gennes (BdG) approach when pairing is included. This was used to study fermionic atoms with resonant interactions in an optical lattice \cite{Watanabe2008}, where the pairing gap is comparable to the Fermi energy, while for neutrons it is smaller by an order of magnitude.
We shall take the pairing interaction $g$ to be a constant, but to allow for the effects of correlations between particles and the momentum dependence of the neutron-neutron interaction, we shall choose the strength of the pairing interaction so that it gives pairing gaps for uniform neutron matter in accord with the results of detailed microscopic calculations.
When the momentum of the superfluid is nonzero,  the single-particle states in the BdG  approximation satisfy the equation \cite{Watanabe2008}
\be
\left(
\begin{array}{cc}
\tilde H_{\QQ}'(\rr)&\tilde \Delta(\rr)\\
\tilde \Delta^*(\rr)&-\tilde H_{-\QQ}'(\rr)
\end{array} \right)
\left(
\begin{array}{c}
\tilde u_i(\rr)\\
\tilde v_i(\rr)
\end{array} \right)
=\epsilon_i \left(
\begin{array}{c}
\tilde u_i(\rr)\\
\tilde v_i(\rr)
\end{array} \right).  
\ee
 Here, $\QQ$ is the momentum per particle in the condensate \cite{hbar},
 \be
\tilde H_{\QQ}'(\rr) =   \frac{1}{2m}   (-\rmi \bm\nabla+\QQ+\kk)^2   +V_{\rm ext}(\rr) - \mu,
 \ee
\be
\tilde\Delta(\rr)=-g\sum_i {\tilde u}_i(\rr) {\tilde v}^*_i(\rr),
\ee
 and the state index $i$ is shorthand for the pseudomomentum $\kk$ (which lies within the first Brillouin zone) and a band index.  The functions
 $\tilde u_i(\rr)=u_i(\rr)\rme^{-\rmi (\QQ+\kk)\cdot \rr}$ and $\tilde v_i(\rr)=v_i(\rr)\rme^{\rmi (\QQ-\kk)\cdot \rr}$ have the same periodicity as the lattice.
The superfluid number density tensor is calculated from the energy density $\mathcal E$ via the expression
\be
 n_{ij}^s= m\frac{\partial^2  {\mathcal E}(n, \QQ)}{\partial Q_i  \partial Q_j}.
\ee
Details of the calculational methods are given in Ref.~\cite{Watanabe2008}.  We take the periodic potential to be one dimensional, 
\be
V_{\rm ext}(\rr)=V_K (\rme^{\rmi Kz}+\rme^{-\rmi Kz}),
\ee
where $V_K$ is real.  In terms of the variables used in Ref.~\cite{Watanabe2008}, $|V_K|=s E_R/4$, where $E_R=K^2/8m=\pi^2/(2md^2)$, $d$ is the lattice period, and $s$ is a dimensionless measure of the strength of the lattice potential.
With this potential, $n^s_{zz} \neq n$, where $n$ is the average particle density, but it follows from Galilean invariance in the $x$ and $y$ directions that $n^s_{xx}=n^s_{yy}=n$.  

We remark that close to the inner boundary of the crust, it is predicted that ``pasta'' phases with rodlike or platelike nuclei can occur \cite{RavenhallCJPWilson, Hashimoto}.  For these, the superfluid density will be anisotropic.  The calculations for a one-dimensional lattice may thus be regarded as a first approximation for the platelike phase, ``lasagna''.

Figure \ref{Overview} shows that, for weak pairing, the superfluid density is suppressed significantly, but that with increasing strength of the pairing, characterized by $\Delta_{}$, the pairing gap in a uniform medium at the same density, the suppression is strongly reduced.  The physical significance of the results is most simply brought out by plotting $n^s_{zz}/n$ as a function of $\Delta_{}/|V_K|$, Fig.\,\ref{Delta_over_V} \cite{footnote}.  The results are well fitted by the expression 
\be
n^s_{zz}(\Delta_{})=n-\frac{n-n^s_{zz}(0)}{\left[ 1+(\Delta_{}/|V_K|)^2 \right]^{1/2}}\, .
\label{nsvsDelta}
\ee
Thus the effects of band structure on the superfluid density are suppressed dramatically for $\Delta_{} \gtrsim |V_K|$.
\begin{figure}
\includegraphics[width=3.5in]{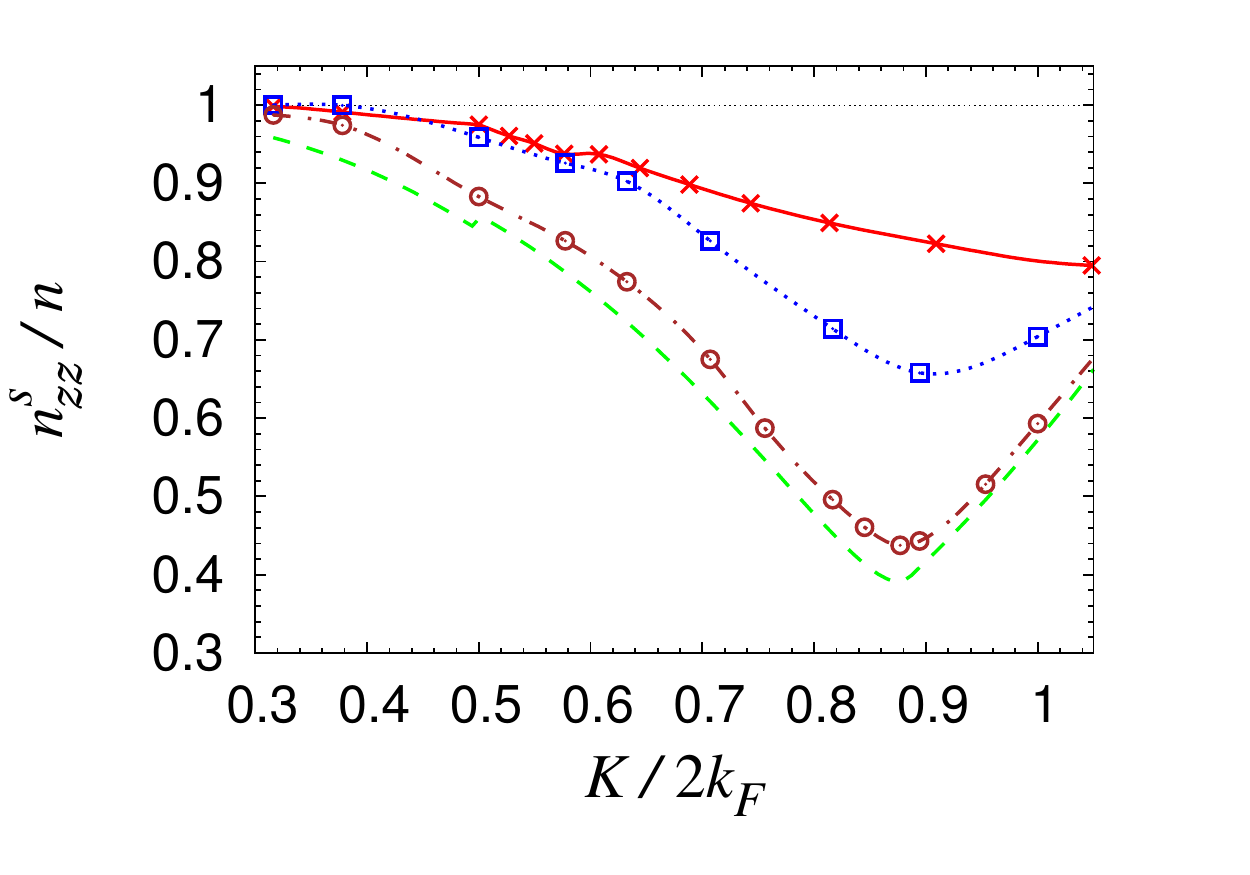}
\caption{Superfluid number density in terms of the total number density as a function of $K/2k_F$, where $k_F$ is the Fermi wave number of the uniform noninteracting Fermi gas at the same density.  The curves are for a lattice potential $V_K=0.25 (K/2k_F)^2 E_F$ with $E_F=k_F^2/2m$. The lines are for the case of  $\Delta_{} \to 0$ (green dashed line), $\Delta_{}\approx  0.0464 E_F$ (brown dashed-dotted line), $\Delta_{}\approx 0.208 E_F$ (blue dotted line), and $\Delta_{}\approx 0.686  E_F$ (red solid line).}
\label{Overview}
\end{figure}
\begin{figure}
\includegraphics[width=3.5in]{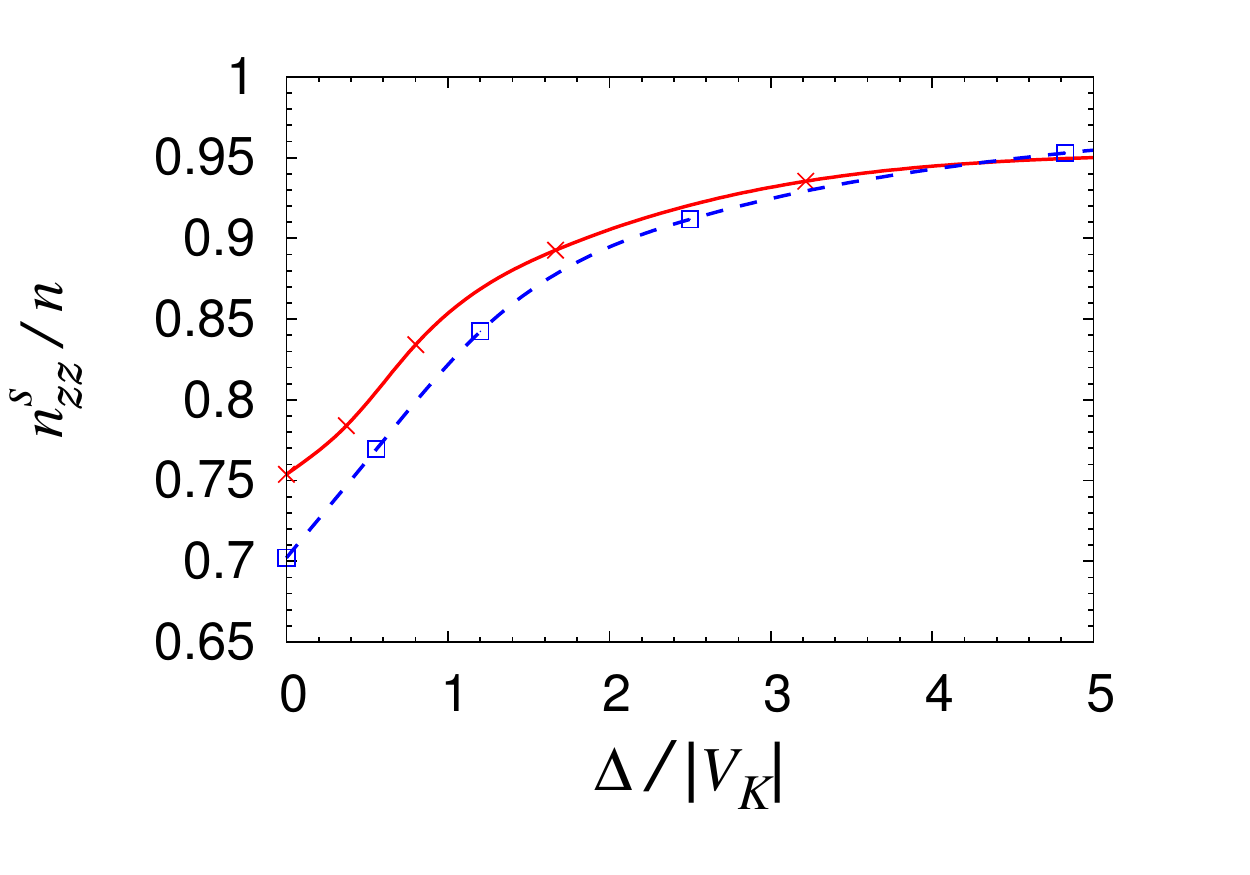}
\caption{Superfluid density $n_{zz}^s/n$ as a function of $\Delta_{}/|V_K|$ for the same lattice potential as in Fig.\,\ref{Overview}.   The red solid line is for $K=2k_F$  and the blue dashed line is for $K=\sqrt{8/3}\,k_F\approx 1.633 \, k_F$. }
\label{Delta_over_V}
\end{figure}

\textit{Application to neutron star crusts.---}    For total nucleon densities in the range of $0.001$--$0.05$~fm$^{-3}$, the calculations of Ref.~\cite{Chamel2012} predict a reduction of the neutron superfluid density by a factor $\sim 3$ or greater, and we shall focus on a density of $0.03$~fm$^{-3}$, where the effect is largest.  The density of neutrons outside nuclei, $n_n^o$, is $\approx 0.024$~fm$^{-3}$, corresponding to a neutron Fermi wave number $k_n=(3\pi^2n_n^o)^{1/3}$ of  $0.89$~fm$^{-1}$ and a Fermi energy $k_n^2/2m\approx 16.4$~MeV.  These densities are so high that analytical results for uniform matter at low densities cannot be applied, and there are many estimates of gaps based on many-body theory.  Most calculations predict that the $^1$S$_0$ gap is close to its maximum value at these densities  \cite{compilationofgaps}.  Calculations in which the pairing interaction is taken to be the free-space neutron-neutron interaction predict gaps of approximately $3$~MeV.  A variety of theoretical approaches lead to the conclusion that gaps will be suppressed due to the effect of the medium on the pairing interaction, primarily exchange of spin fluctuations, and numerically this is typically a factor of $2$--$3$ \cite{Induced_interactions}.   Thus, one may expect gaps to lie in the range of $1$--$1.5$~MeV.  

What makes calculations for crustal matter challenging is the fact that the lattice potential has many Fourier components, with wave vectors corresponding to the reciprocal lattice vectors (RLVs) of the solid, rather than the single pair of them considered in the BdG calculations above.  The large number of Fourier components is reflected in the high number of occupied bands.  RLVs with $K_i\ge2k_n$ have little effect because scattering of fermions between two states in the vicinity of the Fermi surface is impossible for such wave vectors.   However, although the lattice potential has many Fourier components,  they are relatively weak, in the sense that their magnitudes are much less than the Fermi energy.   This is seen from results for  $V_K$ obtained from the Fourier transform of the potential for a single spherical cell surrounding a nucleus obtained by Pearson {\it et al.} \cite{Pearson2015,Chamel_pc}, which are shown in 
Fig.~\ref{Fouriertransform}.  The magnitude of the lattice potential has a maximum value of $2$~MeV for $K=0$ and falls rapidly with $K$: for wave numbers greater than about $0.3 k_n$ it is comparable to estimates of the superfluid gap, and for $K\gtrsim 0.5 k_n$ it is less than one tenth of the superfluid gap.

\begin{figure}
\includegraphics[width=3.5in]{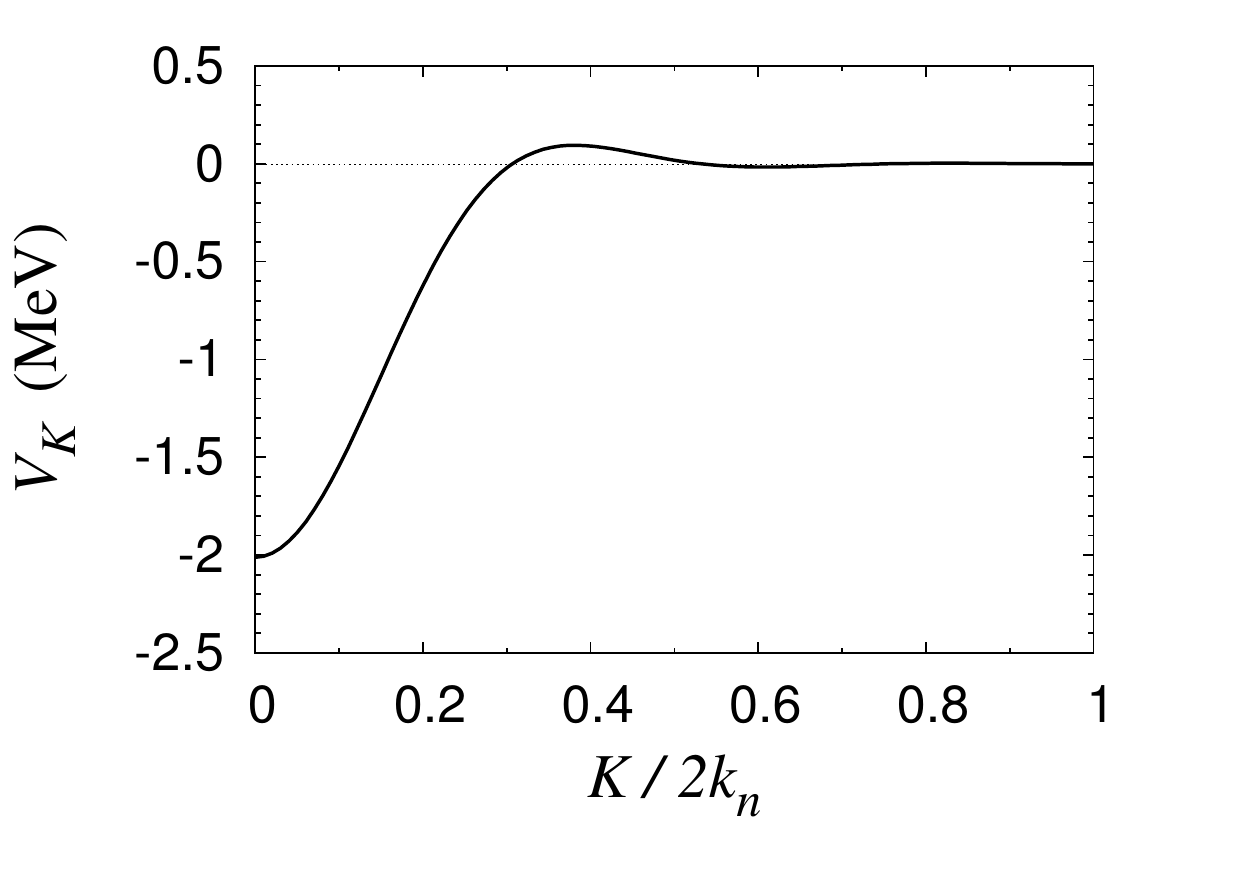}
\caption{Lattice potential $V_K$ as a function of $K/2k_n$ from the work of Ref.\ \cite{Pearson2015}.  Because there are so many reciprocal lattice vectors with $K<2k_n$, $K$ is treated as a continuous variable.}
\label{Fouriertransform}
\end{figure}

To estimate quantitatively the effect of many RLVs, we assume that pairs of RLVs $\KK_i$ and $-\KK_i$ contribute independently to the superfluid density.  This should be a reasonable first approximation because processes involving more than one pair of RLVs are of higher order in the Fourier components of the lattice potential, which are small in magnitude compared with the Fermi energy.
For a single pair of RLVs, the superfluid density tensor for arbitrary orientations of $\KK$ is given by
\be 
n_{ij}^s({\bf K})=n\delta_{ij}- \left[n-n_{zz}^s(K)\right]{\hat K}_i{\hat K}_j.
\ee
When averaged over possible orientations of $\KK$ for cubic symmetry, one finds the superfluid density is a scalar, and given by
\be
\frac{n^s}{n}=1- \frac{1}{3}\left(1-\frac{n_{zz}^s(K, V_K, \Delta)}{n}\right).
\ee 
Thus, the effect of many RLVs in the neutron star crust is given by
\bea
\frac{n^s}{n_n^o}={\prod_{\KK_i}}'
\left\{1- \frac13\left[1-\frac{n_{zz}^s(K, V_K, \Delta)}{n}\right]\right\},
\label{prod}
\eea 
where the prime on the product indicates that, to avoid double counting, only one of the RLVs $\KK_i$ and $-\KK_i$ is to be included.  Here, the quantity ${n_{zz}^s(K, V_K, \Delta)}/{n}$ on the right-hand side is that calculated above from the BdG equations for a sinusoidal potential.  By using the quantity $n_n^o$ as the reference superfluid density on the left side of this equation we have neglected the effect of deeply bound neutron states that contribute little to the neutron density outside nuclei: these occupy filled bands well below the Fermi surface and therefore do not contribute to the neutron current.
Since the reduction of the superfluid density by a single pair of RLVs is typically less than one percent, we may write
\be
\frac{n^s}{n_n^o}\approx \exp\left\{-\frac16\sum_{\KK_i}\left[1-\frac{n_{zz}^s(K, V_K, \Delta)}{n}\right]\right\},
\label{prod_exp}
\ee 
where the sum is over all reciprocal lattice vectors.

On replacing the sum in Eq.~(\ref{prod_exp}) by an integral one finds 
\be
\frac{n^s}{n_n^o}\approx\exp{\left\{ - \frac{2n_n^o}{n_N}\int_0^1 x^2d x\left[ 1-\frac{n_{zz}^s(K, V_K, \Delta )}{n} \right] \right\}},
\label{prod_exp_int}
\ee
where $x=K/(2k_n)$ and $n_N$ is the density of nuclei.

Solving the BdG equations for the range of values of $V_K$ and $K$ encountered in the integral in Eq.~(\ref{prod_exp_int}) is time consuming, so we adopt a simplified approach to estimate the effects of band structure with and without pairing.  First, we assume that in the presence of a superfluid gap, $n_{zz}^s$ scales as in Eq.~(\ref{nsvsDelta}).  We have calculated \mbox{$n_{zz}^s(K, V_K, \Delta =0)$}  for a range of different $|V_K|$ and $K$ values.  The function 
$1-{n_{zz}^s(K, V_K, \Delta =0)}/{n} \approx (1+3.5 x){|V_K|}/{E_F^o}$ with $E_F^o = (3\pi^2 n_n^o)^{2/3}/2m$
gives a reasonable first approximation for the values of $|V_K|$ as a function of $K$ for the potential shown in Fig.~\ref{Fouriertransform}.  In the limit of no pairing, this approximation gives $n^s/n_n^o\approx 0.20$.  This represents a factor of 5 reduction, which is considerable, although not as large as the values Chamel found in Ref.~\cite{Chamel2012}, which were more than a factor of 10.  
What is particularly interesting are the results when pairing is included: we find $n^s/n_n^o\approx 0.64$, a $36$\% reduction for a gap of $1$~MeV, and  $n^s/n_n^o\approx 0.71$, only a $29$\% reduction, for a gap of $1.5$~MeV.

Our calculations demonstrate that  pairing greatly suppresses the effects of band structure because  the magnitude of the periodic potential $|V_K|$ is considerably less than the pairing gap $\Delta$ for the vast majority of  reciprocal lattice vectors.  Calculations that treat better the effects of the many Fourier components of the lattice potential need to be done, but it should be enough to consider only a limited number of components  with wave numbers  $\lesssim 0.5k_n$.  Our calculations suggest a reduction of the neutron superfluid density by some tens of a percent when pairing is included, as opposed to the 1 order of magnitude predicted in the limit of a small gap.   We therefore conclude that the effects of band structure on the neutron superfluid density are modest when pairing is taken into account and, consequently, that glitch models based on the superfluid density of neutrons in the inner crust are still tenable.

Understanding the mechanism for pinning vortices in the neutron superfluid to the lattice is an important challenge for modeling the time evolution of neutron star glitches, but the arguments in this Letter do not depend on it.

Finally, we note that cold atomic gases in optical lattices are a useful system for investigating experimentally the suppression of band structure effects by pairing, since the strengths of the periodic potential and of the pairing interaction can both be varied.  Such experiments could provide confirmation of the finding of this Letter that band structure effects are suppressed if the strength of the periodic potential is less than the pairing gap, and  it is not necessary that the pairing gap be comparable to the Fermi energy.
 
We are very grateful to Nicolas Chamel for correspondence and for sharing with us unpublished results from the work of Ref.~\cite{Pearson2015}. Discussions on neutron pairing gaps with Alex Gezerlis were useful. C.~J.~P. is grateful to Ole Krogh Andersen, Mark Rudner, Michael Schecter and Hans Skriver for discussions during the early stages of this work.   G.~W. thanks Qijin Chen for discussions. He was supported by the Zhejiang University 100 Plan, by the Junior 1000 Talents Plan of China, and by NSF of China (Grant No. 11674283).  The bulk of the numerical calculations were performed on the IBS-PCS cluster ``Fermi'' (NFEC-2016-08-211102).  Nordita provided travel and subsistence support that made possible the completion of this work.  This work was also supported in part by NewCompStar, COST Action MP1304.


\begin{thebibliography}{99}
\bibitem{footnote} The corresponding plots in terms of the average pairing gap in the presence of the lattice (rather than $\Delta_{}$) are essentially indistinguishable from Fig.~\ref{Delta_over_V}.
\end{thebibliography}
\end{document}